\documentclass[doublecol]{epl2} 
\usepackage{amssymb}
\usepackage{bm}
\usepackage{epsfig}
\usepackage{wasysym}

\usepackage{xcolor}

\title{Unifying topological phase transitions in noninteracting, interacting, and periodically driven systems}

\author{Paolo Molignini\inst{1} \and R. Chitra\inst{1} \and Wei Chen\inst{1,2}}
\shortauthor{P. Molignini \etal}

\institute{                    
  \inst{1} Institute for Theoretical Physics, ETH Z\"{u}rich, 8093 Zurich, Switzerland\\
  \inst{2} Department of Physics, PUC-Rio, 22451-900 Rio de Janeiro, Brazil
}

\pacs{68.35.Rh}{Phase transitions and critical phenomena}
\pacs{64.60.F-}{Equilibrium properties near critical points, critical exponents}
\pacs{64.60.ae}{Renormalization-group theory}

\abstract{
Topological phase transitions %track changes in topological properties of a system and 
occur in real materials as well as quantum engineered systems, all of which differ greatly in terms of dimensionality, symmetries, interactions, and driving, and hence require a variety of techniques and concepts to describe their topological properties.
For instance, %depending on the system, 
topology may be accessed from single-particle Bloch wave functions, Green's functions, or many-body wave functions.
We demonstrate that despite this diversity, all topological phase transitions display a universal feature:  namely, a divergence of the curvature function that composes the topological invariant at the critical point.
This feature can be exploited via a renormalization-group-like methodology to describe topological phase transitions.
This approach serves to extend notions of correlation function, critical exponents, scaling laws and universality classes used in Landau theory to characterize topological phase transitions in a unified manner.
}

\begin{document}

\maketitle

\section{Introduction}
%Following the discovery of the fractional quantum Hall effect~\cite{Tsui:1982, Laughlin:1983}, topological order representing radically new phases of matter has revolutionized the field of condensed matter physics~\cite{Wen:1990, HassanReview:2010, Bernevig13}.
%A plethora of exotic topological orders, ranging from Majorana phases in spinless fermionic systems~\cite{Kitaev2001}, topological insulators and superconductors~\cite{Kane-Mele2005, Kane-Mele2005-2, Hsieh:2008}, to Weyl semimetals~\cite{Wan:2011,Xu:2016} have been identified.
%Contrary to the traditional Landau formalism based on spontaneous symmetry breaking~\cite{Landau,Miransky-book}, topological phase transitions (TPTs) in these systems are not described by local order parameters but by robust ground state degeneracies, long-range quantum entanglement~\cite{Wen:1995, XieChen:2010,Fidkowski:2010} and quantized geometric phases~\cite{Thouless:1982, Wen:1989, Qi:2008}.  
Topological phase transitions (TPTs) rely on discrete changes in an integer topological invariant ${\cal C}$ and understanding them necessitates an extension of the paradigm of quantum phase transitions\cite{Sachdev-book,Continentino-book}.  
Upon tuning one or multiple system parameters ${\bf M}=(M_{1},M_{2}...)$, 
%such as a magnetic field\cite{Oreg10,Mourik12} or a time-periodic potential~\cite{Rechtsman13,Peng16}, 
$\cal C$ jumps abruptly from one integer to another at the critical point ${\bf M}_{c}$. 
Drawing an analogy to the usual Landau second order phase transitions, TPTs usually involve a gap closure in the single-particle or many-body energy spectrum. 
Despite this similarity, the absence of a local Landau order parameter
renders the notion of critical behavior rather ambiguous at a glance.
%Additionally, since the topological invariant ${\cal C}$ remains an integer in every phase and only jumps discretely at the TPTs, it is unclear how quantum criticality based on diverging susceptibilities.

In recent years, there have been several attempts at describing the criticality of TPTs, most notably through the behavior of entanglement measures such as the entanglement entropy (EE) and the entanglement spectrum (ES)~\cite{Levin:2006, Kitaev:2006, Thomale:2010, Sterdyniak:2011, Jiang:2012,Hsieh:2014, Cho:2017, Zhang:2018}. 
%The ES is in direct correspondence with the excitation spectrum of the physical edge modes, while the EE can exhibit non-analiticities in concomitance with a TPT, from which a scaling behavior could in principle be extracted. Nevertheless, these signatures do not appear to be universal and have be shown to be sometimes attributable to other phenomena~\cite{Cho:2017,Yu:2009}.
Alternate formulations proposed include universal central charges in CFT formulations of one dimensional systems~\cite{Yates:2018,Berdanier:2018}, topological defect generation~\cite{Liou:2018, Manna:2019}, real-space topological markers~\cite{Bianco:2011,Caio:2019}, and derivatives of suitably defined thermodynamic potentials~\cite{Kempkes:2016, Quelle:2016, Cats:2018}.
Though these approaches brought forward fundamental advances in understanding topological criticality,  they have not yet offered a generalized and thorough correspondence to the usual concepts of Landau quantum criticality so far.

Recently, a framework based on scaling theory was proposed to discuss the criticality of TPTs \cite{Chen:2016,Chen-Sigrist:2016,Chen:2017,Chen-Sigrist-book:2019}. 
The theory relies on the topological invariant ${\cal C}$ being generally an integration of a certain ``curvature function" $F$ over the momentum or flux space.
The curvature function depends on the dimensionality and symmetry class of the system\cite{Chen-Schnyder:2019}, whether it is static or periodically driven\cite{Molignini:2018,Molignini:2019}, and whether it is interacting or not\cite{Kourtis:2017,Chen:2018}.
A vast majority of topological materials display a universal feature in $F$: as the system approaches the TPT where the gap closes, $F$ gradually diverges, with the divergence changing sign across the transition.
This feature allows the classification of TPTs according to standard concepts of critical exponents and universality classes.
Furthermore, due to the conservation of the topological invariant, scaling laws linking the exponents  naturally emerge~\cite{Chen:2017,Chen-Sigrist-book:2019,Chen-Schnyder:2019}.
Based on this notion of divergence, a simple renormalization group (RG) approach is proposed to analyze TPTs.
This procedure is a demonstrably efficient tool to pinpoint and characterize TPTs in multi-dimensional parameter spaces\cite{Chen:2016,Chen-Sigrist:2016}, a typically cumbersome task in both interacting and periodically driven systems\cite{Molignini:2018,Molignini:2019,Chen:2018,Kourtis:2017}.
%To our knowledge, this is the only known unified scheme that is applicable to such a broad variety of topological systems.

In this review, we outline this unified description of TPTs and its applications to a variety of static, periodically driven, weakly and strongly interacting systems.
We mainly focus on one (1D) and two-dimensional (2D) Dirac models, although higher dimensional cases and more exotic topology have also been discussed~\cite{Chen:2016,Chen-Sigrist:2016,Chen:2017,Kourtis:2017,Molignini:2018,Chen:2018,Molignini:2019,Chen-Sigrist-book:2019,Chen-Schnyder:2019}.
%We demonstrate several surprising results uncovered by this method, such as quantum multi-criticality induced by periodic driving.
%%%%%%%%%%%%%%%%%%%%%%
\begin{figure}[ht]
\centering
\includegraphics[width=0.99\columnwidth]{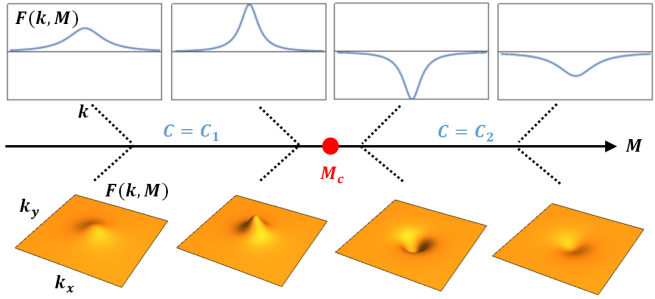}
\caption{Schematics of the divergence of the curvature function at the critical point ${\bf M}_{c}$ in 1D (top) and 2D (bottom) systems. The height and width of the Lorentzian peak are $F({\bf k}_{0},{\bf M})$ and $1/\xi$, repectively, and have critical exponents $\gamma$ and $\nu$. }
\label{fig:curv_fn_divergence}
\end{figure}
%%%%%%%%%%%%%%%%%%%%%%

\section{Correlation functions, critical exponents, and universality classes}
We consider a topological system whose topology is determined by a set of tunable parameters ${\bf M}=\left(M_{1},M_{2},...\right)$ in the Hamiltonian. 
Different topological phases correspond to different quantized values of a topological invariant $\mathcal{C}$ constructed from integrating a curvature function $F(\mathbf{k}, {\bf M})$ over the Brillouin zone (BZ)
\begin{eqnarray}
\mathcal{C}=\int \mathrm{d}^D k \: F(\mathbf{k}, {\bf M})\;.
\label{topological_invariant_integration}
\end{eqnarray}
The curvature function in the vicinity of a high symmetry point (HSP) ${\bf k}_{0}$ typically displays a Lorentzian shape
\begin{eqnarray}
F({\bf k}_{0}+\delta{\bf k},{\bf M})=\frac{F({\bf k}_{0},{\bf M})}{1+\xi^{2}\delta k^{2}}\;,
\label{Lorentzian_shape}
\end{eqnarray}
that reflects the evenness of the curvature function $F({\bf k}_{0}+\delta{\bf k},{\bf M})=F({\bf k}_{0}-\delta{\bf k},{\bf M})$ around the HSP, where $1/\xi$ defines the width of the multidimensional peak.
Approaching the critical point ${\bf M}\rightarrow{\bf M}_{c}$ the peak gradually diverges, flipping sign across the transition\cite{Chen:2017,Chen-Sigrist-book:2019,Chen-Schnyder:2019}
\begin{eqnarray}
&&\lim_{{\bf M}\rightarrow {\bf M}_{c}^{+}}F({\bf k}_{0},{\bf M})=-\lim_{{\bf M}\rightarrow {\bf M}_{c}^{-}}F({\bf k}_{0},{\bf M})=\pm\infty\;,
\nonumber \\
&&\lim_{{\bf M}\rightarrow {\bf M}_{c}}\xi=\infty\;,
\label{F0_Xi_divergence}
\end{eqnarray}
This is represented schematically in Fig.~\ref{fig:curv_fn_divergence}.

This divergence helps us introduce the critical exponents $\gamma,\nu$ of the TPT: 
\begin{eqnarray}
|F({\bf k}_{0},{\bf M})|\propto|{\bf M}-{\bf M}_{c}|^{-\gamma}\;,\;\;\;\xi\propto|{\bf M}-{\bf M}_{c}|^{-\nu}\;,
\label{F0_xi_critical_exponents}
\end{eqnarray}
The conservation of the topological invariant ${\cal C}=\mathrm{const.}$ as the transition is
approached from one side or the other,
\begin{equation}
{\cal C}=F({\bf k}_{0},{\bf M})\left(\prod_{i=1}^{D}\int_{-\xi^{-1}}^{\xi^{-1}}\frac{dk_{i}}{1+\xi^{2}k_{i}^{2}}\right)
\nonumber  \propto
\frac{F({\bf k}_{0},{\bf M})}{\xi^{D}},
\label{Cdiv_peak_divergent}
\end{equation}
yields a scaling law that constraints the exponents\cite{Chen:2017,Chen-Sigrist-book:2019,Chen-Schnyder:2019}
\begin{eqnarray}
\gamma=D\nu\;,
\label{scaling_law_peak_divergence}
\end{eqnarray}
where $D$ is the dimensionality of the problem.
These exponents serve to classify TPTs into different universality classes, regardless of the details of the system,
%whether the system is noninteracting, interacting, or periodically driven, 
as we will demonstrate in the following sections.

Analogous  to the Landau formalism, we can introduce a correlation function that characterizes the TPT.
First, we calculate the Fourier transform of the curvature function
\begin{eqnarray}
\tilde{F}({\bf R})=\int \frac{d^{D}{\bf k}}{(2\pi)^{D}}\;e^{i{\bf k}\cdot{\bf R}}\;F({\bf k},M)\;.
\label{general_correlation_function}
\end{eqnarray}
and define the Wannier state  
\begin{eqnarray}
|{\bf R} n\rangle=\frac{1}{N}\sum_{\bf k}e^{i {\bf k}\cdot({\hat{\bf r}}-{\bf R})}|u_{n{\bf k}}\rangle\;,
\end{eqnarray}
where $|u_{n{\bf k}}\rangle$ is the single-particle Bloch state or many-body state from which topology is defined.
With the Wannier wave function $W_{n}({\bf r-R})=\langle{\bf r}|{\bf R}n\rangle$ centered at the home cell {\bf  R}, Eq.~(\ref{general_correlation_function}) indicates that the correlation function $\tilde{F}({\bf R})=\langle{\bf R}|{\hat {\bf R}}|{\bf 0}\rangle$ is a measure of the overlap of Wannier functions centered at two home cells that are distance ${\bf R}$ apart\cite{Chen:2017,Chen-Sigrist-book:2019,Chen-Schnyder:2019}.
Combining the Lorentzian form in Eq.~(\ref{Lorentzian_shape}) with the Fourier transform in Eq.~(\ref{general_correlation_function}), we see that the correlation function decays with correlation length $\xi$. 
The divergence of $\xi$ described by Eq.~(\ref{F0_Xi_divergence}) then indicates that, near the TPT, the Wannier functions become relatively extended and  have large overlaps. 
This phenomenon can in turn be interpreted as the  equivalence of scale invariance at the critical point seen in standard phase transitions.

\section{The curvature renormalization group approach}
The divergent behavior of the curvature function suggests the construction of an iterative procedure to search for the trajectory in the parameter space (RG flow) along which the divergence is  mitigated but the topology remains unchanged~\cite{Chen:2016, Chen-Sigrist:2016, Chen:2017}.
Under this procedure, the system will gradually move away from the critical points of the flow.
%representing the transitions between different topological phases.
By mapping out these critical points, the topological phase diagram can thus be identified. 
The iterative procedure demands that at a given parameter set ${\bf M}$, one searches for a new parameter ${\bf M}^{\prime}$ that satisfies
\begin{equation}
F({\bf k}_0, {\bf M}^{\prime}) = F({\bf k}_0 + \delta {\bf k}, {\bf M}),
\label{cond-RG-eqn}
\end{equation}
where ${\bf k}_{0}$ is a HSP and $\delta {\bf k}$ is a small deviation away from it. 
The divergence of $F({\bf k}_{0},{\bf M})$ is gradually reduced under this procedure, as can be rigorously proved by analyzing it in Fourier space~\cite{Chen:2016}. 
Because it renormalizes the profile of the curvature function, this method has been referred to as the curvature renormalization group (CRG) approach.

Writing $\mathrm{d}M_{i} = M_{i}^{\prime} - M_{i}$ and $\delta k_j^2 \equiv \mathrm{d}l$, and expanding Eq.~(\ref{cond-RG-eqn}) to leading order yields the generic RG equation
\begin{equation}
\frac{\mathrm{d}M_{i}}{\mathrm{d} l} = \frac{1}{2} \frac{\partial^2_k F({\bf k}, {\bf M}) \big|_{k=k_0}}{\partial_{M_{i}} F({\bf k}_{0}, {\bf M})},
\label{generic_RG_equation}
\end{equation}
from which the RG flow can be obtained.
The RG flow contains both critical and fixed points~\cite{Kourtis:2017}:
\begin{eqnarray}
{\rm Critical\;point}:&&\left|\frac{d{\bf M}}{dl}\right|\rightarrow\infty, %\;{\rm flow\;directs\;away},
\nonumber \\
{\rm \;Fixed\;point}:&&\left|\frac{d{\bf M}}{dl}\right|\rightarrow 0. %\;{\rm flow\;directs\;into},
\nonumber
\label{identifying_Mc_Mf}
\end{eqnarray}

In non-interacting systems, obtaining analytic expressions for the critical points is often easy.
However, to see the full flow pattern, one resorts to numerics. 
For this purpose, one can implement Eq.~(\ref{generic_RG_equation}) on a discrete mesh:
\begin{equation}
\frac{\mathrm{d}M_{i}}{\mathrm{d} l} = \frac{\Delta M_{i}}{(\Delta {\bf k})^{2}}\frac{F({\bf k}_{0}+\Delta {\bf k},{\bf M})-F({\bf k}_{0},{\bf M})}{F({\bf k}_{0},{\bf M}+\Delta M_{i}{\hat{\bf M}}_{i})-F({\bf k}_{0},{\bf M})},
\label{generic_RG_equation_discrete}
\end{equation}
where $\Delta {\bf k}$ is the grid spacing in the momentum space,  and $\Delta M_{i}$ is the spacing in  parameter space along  the unit vector ${\hat{\bf M}}_{i}$. 
Eq.~(\ref{generic_RG_equation_discrete}) indicates that at a given ${\bf M}$, we only require the knowledge of the curvature function at three points: $F({\bf k}_{0}+\Delta {\bf k},{\bf M})$, $F({\bf k}_{0},{\bf M})$, and $F({\bf k}_{0},{\bf M}+\Delta M_{i}{\hat{\bf M}}_{i})$  to obtain the RG flow $\mathrm{d}M_{i}/\mathrm{d}l$ in Eq.~(\ref{generic_RG_equation_discrete}) without the need for the explicit integration in Eq.~(\ref{topological_invariant_integration}).
Hence, the CRG approach is a very efficient way to determine the TPTs circumventing the direct calculation of topological invariants in a multi-dimensional parameter space.
We now demonstrate these statements in concrete systems.

\section{Noninteracting systems}
We first discuss noninteracting, static systems.
%that serve as the basis for the interacting and periodically driven systems presented in the following sections.
%For conciseness, we limit the discussion to gapped 1D and 2D systems throughout the article. 
The generic form of the Hamiltonian under consideration is of  a Dirac Hamiltonian 
\begin{eqnarray}
H({\bf k})={\bf d}({\bf k})\cdot{\boldsymbol\Gamma}\;,
\label{Dirac_Hamiltonian}
\end{eqnarray}
where the matrices $\Gamma^{a}$ %are matrices that 
satisfy %the Clifford algebra 
$\left\{\Gamma^{a},\Gamma^{b}\right\}=2\delta_{ab}$.
The low-energy dispersion is assumed to be
\begin{eqnarray}
E_{\pm}({\bf k})=\pm\vert {\bf d}(k)\vert=\pm\left(k^{2n}+M^{2}\right)^{1/2}\;,
\label{general_Dirac_dispersion}
\end{eqnarray}
where $M$ is the gap and  the integer $n$ is the order of band crossing at the TPT (a.k.a. the dynamic exponent\cite{Continentino-book}). 
From the dispersion we infer  the only characteristic length scale of the problem $\xi\sim k^{-1}\sim |M|^{1/n}$, which immediately implies that the critical exponent in Eq.~(\ref{F0_xi_critical_exponents}) is $\nu=1/n$, and $\gamma=D/n$ according to the scaling law in Eq.~(\ref{scaling_law_peak_divergence}). 
In other words, the order of band crossing $n$ determines the universality class of the TPT\cite{Chen-Schnyder:2019}.

In a majority of 1D systems, the topological invariant in Eq.~(\ref{topological_invariant_integration}) is the integration of the Berry connection %of the valence band (assuming only one)
\begin{eqnarray}
F(k,{\bf M})=\langle u_{k}|i\partial_{k}|u_{k}\rangle\;,
\label{Berry_connection_1D}
\end{eqnarray}
where $|u_{k}\rangle$ refers to the periodic part of the Bloch state of the valence band. 
The Berry connection takes the Lorentzian shape of Eq.~(\ref{Lorentzian_shape}) in the appropriate gauge and its Fourier transform yields a Wannier state correlation function\cite{Marzari97,Wang06,Marzari12,Gradhand12} 
\begin{eqnarray}
\tilde{F}_{1D}(R)=\langle 0|{\hat r}|R\rangle\;,
\label{1D_class_BDI_correlation_fn}
\end{eqnarray}
that decays exponentially with the correlation length $\xi$. 
1D Dirac models with linear band crossing $n=1$, such as the Su-Schrieffer-Heeger (SSH) model\cite{Su79} and the Kitaev chain\cite{Kitaev2001}, host TPTs with critical exponents $\gamma=\nu=1$ %that satisfy the scaling law in Eq.~(\ref{scaling_law_peak_divergence})\cite{Chen:2017,Molignini:2018}.  
In the case of the Kitaev chain, the Berry connection in Eq.~(\ref{Berry_connection_1D}) is formulated in the Majorana basis, and hence the correlation function in Eq.~(\ref{1D_class_BDI_correlation_fn}) measures the overlap of Majorana-Wannier functions separated by a distance $R$.

For 2D time-reversal breaking systems, the topological invariant is the dimensionless Hall conductance calculated from integrating the gauge-invariant Berry curvature
\begin{eqnarray}
&&F({\bf k},{\bf M})=\partial_{k_{x}}\langle u_{\bf k}|i\partial_{k_{y}}|u_{\bf k}\rangle-\partial_{k_{y}}\langle u_{\bf k}|i\partial_{k_{x}}|u_{\bf k}\rangle
\nonumber \\
&&=\frac{1}{2d^{3}}{\bf d}\cdot\partial_{k_{x}}{\bf d}\times\partial_{k_{y}}{\bf d}=\frac{1}{2}{\hat{\bf d}}\cdot\partial_{k_{x}}{\hat{\bf d}}\times\partial_{k_{y}}{\hat{\bf d}}\;,
\label{Berry_curvature_2D}
\end{eqnarray}
which simply counts the skyrmion number in ${\bf k}-$ space associated with the ${\hat{\bf d}}$-vector in Eq.~(\ref{Dirac_Hamiltonian}). 
The corresponding Wannier state correlation function takes the form\cite{Wang06,Marzari12,Gradhand12}
\begin{eqnarray}
\tilde{F}_{2D}({\bf R})=-i\langle{\bf R}|({\bf R\times{\hat r}})_{z}|{\bf 0}\rangle,
\label{Wannier_correlation_FT_Berry_curvature}
\end{eqnarray} 
and  due to the Lorentzian shape in Eq.~(\ref{Lorentzian_shape}), decays with correlation length $\xi$ for  systems  such as 2D Chern insulators\cite{Bernevig13} with linear band crossings  $n=1$. 
The order of band crossing $n$ in 2D is found to be closely related to non-spatial and crystalline symmetries of the material\cite{Chen-Schnyder:2019}.
For instance, higher order band crossings $n>1$ can be stabilized by a discrete rotational symmetry\cite{Fang12,Yang14}.
The critical exponents $\nu=1/n$ and $\gamma=2/n$ thus imply that the universality class is intimately associated with the crystalline symmetry, akin to the relationship between symmetries and universality classes in the Landau paradigm.
%This is akin to  universality classes  which are  generally determined by the symmetries of the system in the Landau paradigm.

We remark that for the cases where the topological invariant is given by the Pfaffian of the time reversal operator\cite{Chen-Sigrist:2016,Chen-Sigrist-book:2019}, higher order band crossings $n>1$ and higher dimensions\cite{Chen-Schnyder:2019}, the curvature function might display   divergences with more complex structures.
Nevertheless, critical exponents and scaling laws can still be defined and calculated exactly within the CRG formalism.

%%%%%%%%%%%%%%%%%%%%%%%%%%%%%%%%%%%
\begin{figure}[ht]
\centering
\includegraphics[width=0.95\columnwidth]{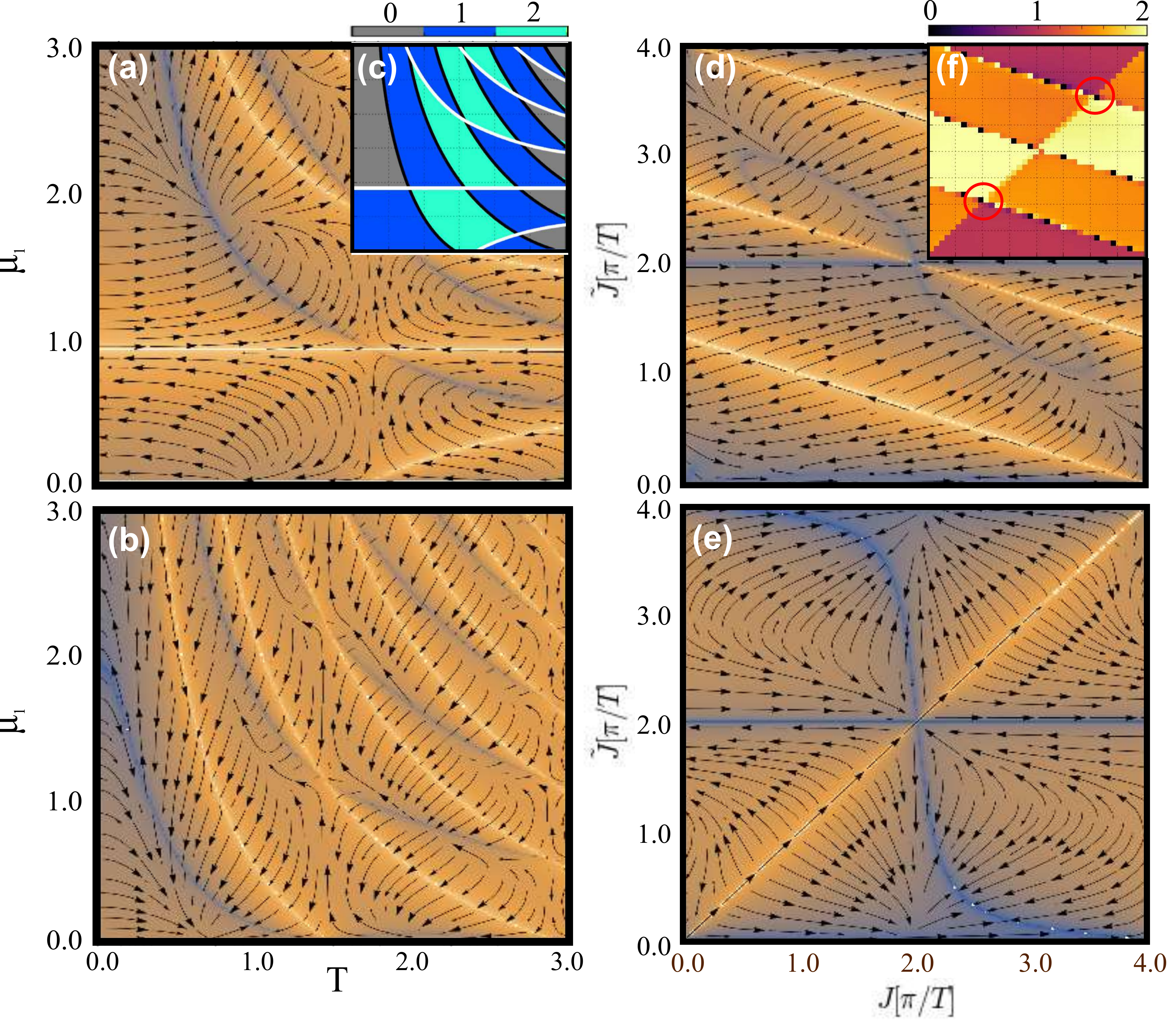}
\caption{CRG flow calculated for the Floquet-Kitaev chain at (a) $k=0$ and (b) $k=\pi$, and for the Floquet-Chern insulator at (d) $\mathbf{k}=(0,0)$ and (e) $\mathbf{k}=(0,-\pi)$. The insets (c) and (e) show the respective topological phase diagrams.}
\label{fig:periodic_driving_results}
\end{figure}
%%%%%%%%%%%%%%%%%%%%%%%%%%%%%%%%%%%

\section{Periodically driven systems}
Periodic driving is %now well known to be 
an important tool to realize %a variety of 
topological order in phases with no inherent topology, \textit{e.g.}in Floquet topological insulators~\cite{Kitagawa,Lindner, Cayssol:2013, Harper:2017, Roy:2017,Esin:2018}, Floquet topological superconductors~\cite{Liu,Thakurathi, Thakurathi:2014, Wang:2014,Sacramento:2015,Thakurathi:2017,Molignini:2017,Molignini:2018,Cadez:2019}, or  Floquet semimetals~\cite{Bucciantini:2017, Huebener:2017, Cao:2017,  ShuChen:2017, ChenZhou:2018, Li:2018}.
Tuning parameters in these driven systems results in nonequilibrium TPTs which can also be characterized by the simple CRG. %as we shall briefly show below

For a general time-periodic system described by the Hamiltonian $H(t)=H(t+T)$, the full dynamics of the system is governed by the time evolution operator
\begin{eqnarray}
U(t) = \mathcal{T} \left\{ \exp \left[ -i \int_0^t \mathrm{d} t' \: H(t') \right] \right\},
\end{eqnarray}
where $\mathcal{T}$ signifies time-ordering.
The operator $U(t)$ accounts for the full time dynamics, including the micromotion between periods.
When $t \to T$, it is usually called Floquet operator and it  induces a discrete quantum map describing stroboscopic dynamics~\cite{Haenggi}. 
We can then define an effective stroboscopic Floquet Hamiltonian via
\begin{eqnarray}
U(T) \equiv e^{-i h_{\rm{eff}} T},
\end{eqnarray}
that contains the full information about the system at multiples of the driving period $T$.  
The eigenvalues of $h_{{\rm eff}} T$  yield the  quasienergy spectrum $\epsilon_{\alpha}$.
The topological phase diagram of the stroboscopic system can be ascertained  by tracking   gap closures and localization  at the  $0$ and $\pi$ quasienergies. Note however, that
though this process yields the correct topography of the phase diagram, it might be necessary to account for micromotion  to obtain the correct topological invariant in a given phase~\cite{Rudner:2013}.

An example of a topological Floquet system in 1D is the periodically driven Kitaev chain with a time-modulated chemical potential $\mu_{0}\rightarrow\mu(t)= \mu_0 + \mu_1 T \sum_{m \in \mathbb{Z}} \delta(t - m T)$~\cite{Thakurathi,Molignini:2018}.
The bulk effective Hamiltonian for this system can be calculated exactly and has a form similar to Eq. (\ref{Dirac_Hamiltonian}), \textit{i.e.} $h_{\mathrm{eff}}(k) \propto d_y(k) \sigma^y + d_z(k) \sigma^z$~\cite{Molignini:2018}, where $\sigma^i$ are Pauli matrices.
The topological invariant can therefore be calculated from the Berry connection of the filled band eigenstate $|u_{k}\rangle$ according to Eq.~(\ref{Berry_connection_1D}) and the Majorana-Wannier state correlation function follows from the construction in Eq.~(\ref{1D_class_BDI_correlation_fn}). 
The CRG approach applied to this problem yields the RG flow in the ${\bf M}=(T,\mu_{1})$ parameter space shown in Figs.~\ref{fig:periodic_driving_results} (a-b), where TPTs appear as critical lines of the CRG flows. 
%Not surprisingly, 
We find that all TPTs in this phase diagram have the critical exponents  $\gamma=\nu=1$, satisfying the scaling law in Eq.~(\ref{scaling_law_peak_divergence}). 
This indicates that the TPTs in both the  static and driven  Kitaev models belong to the same Dirac universality class, despite the latter exhibiting a considerably more complex phase diagram.

In 2D, a  well known  example of a topological Floquet system is a Chern insulator on a square lattice with modulated nearest-neighbor hoppings ${\bf M}=(J,\tilde{J})$ ~\cite{Rudner:2013, Mukherjee:2017,Molignini:2019}.
In this case, the Chern number of the form of Eq.~(\ref{topological_invariant_integration}) is calculated from the Berry \textit{curvature} in Eq.~(\ref{Berry_curvature_2D}) of the stroboscopic bulk effective Hamiltonian $h_{{\rm eff}}(\mathbf{k}) = \mathbf{d}(\mathbf{k}) \cdot \boldsymbol{\sigma}$~\cite{Molignini:2018}.
Fig.~\ref{fig:periodic_driving_results}(d-e) show the CRG flows calculated at the HSPs $(0,0)$ and $(0,-\pi)$, once again revealing all the TPTs.
A detailed analysis reveals a rather unconventional criticality at the TPT $\tilde{J}=J$, with a quadratic gap closure along the nodal loop $k_y = \pm \pi \pm k_x$. 
Here the Lorentzian shapes Eq.~(\ref{Lorentzian_shape}) are deformed to follow the form of the nodal loop.
Nevertheless, we can still extract a set of critical exponents $\gamma  =   \frac{3}{2}$,  $\nu_{\tilde{x}} = \frac{1}{2}$, and $\nu_{y} = 1$, satisfying the scaling law $\nu_{\tilde{x}} + \nu_{y} = \gamma$.
%In contrast to the Lorentzian shapes Eq.~(\ref{Lorentzian_shape}) of usual linear Dirac models, here the curvature function follows the form of the nodal loop and close to the TPT is characterized by a pair of boomerang peaks symmetrically shifted away from the HSPs in the $k_x$ or $k_y$ direction. 
%In this case, we  extract a set of critical exponents from the divergences of the boomerang shape: $\gamma  =   \frac{3}{2}$,  $\nu_{\tilde{x}} = \frac{1}{2}$, and $\nu_{y} = 1$, satisfying the scaling law $\nu_{\tilde{x}} + \nu_{y} = \gamma$.
These different numerical values show that these TPTs in the Floquet problem belong to a new universality class different from the Dirac classes discussed earlier. 
This new universality class stems from a new emergent symmetry of the driven model.
Additionally, this Floquet-Chern insulator also exhibits multi-critical points (red circles in Fig.~\ref{fig:periodic_driving_results}(f)) where linear Dirac-like and nodal loop-like features coexist\cite{Molignini:2019}. 
%These peculiar characteristics exemplify periodic driving as a promising tool to fabricate exotic quantum phases, as well as quantum multi-criticality in topological systems\cite{Tupitsyn10,Rufo19}. 

\begin{figure}[ht]
\centering
\includegraphics[width=0.99\columnwidth]{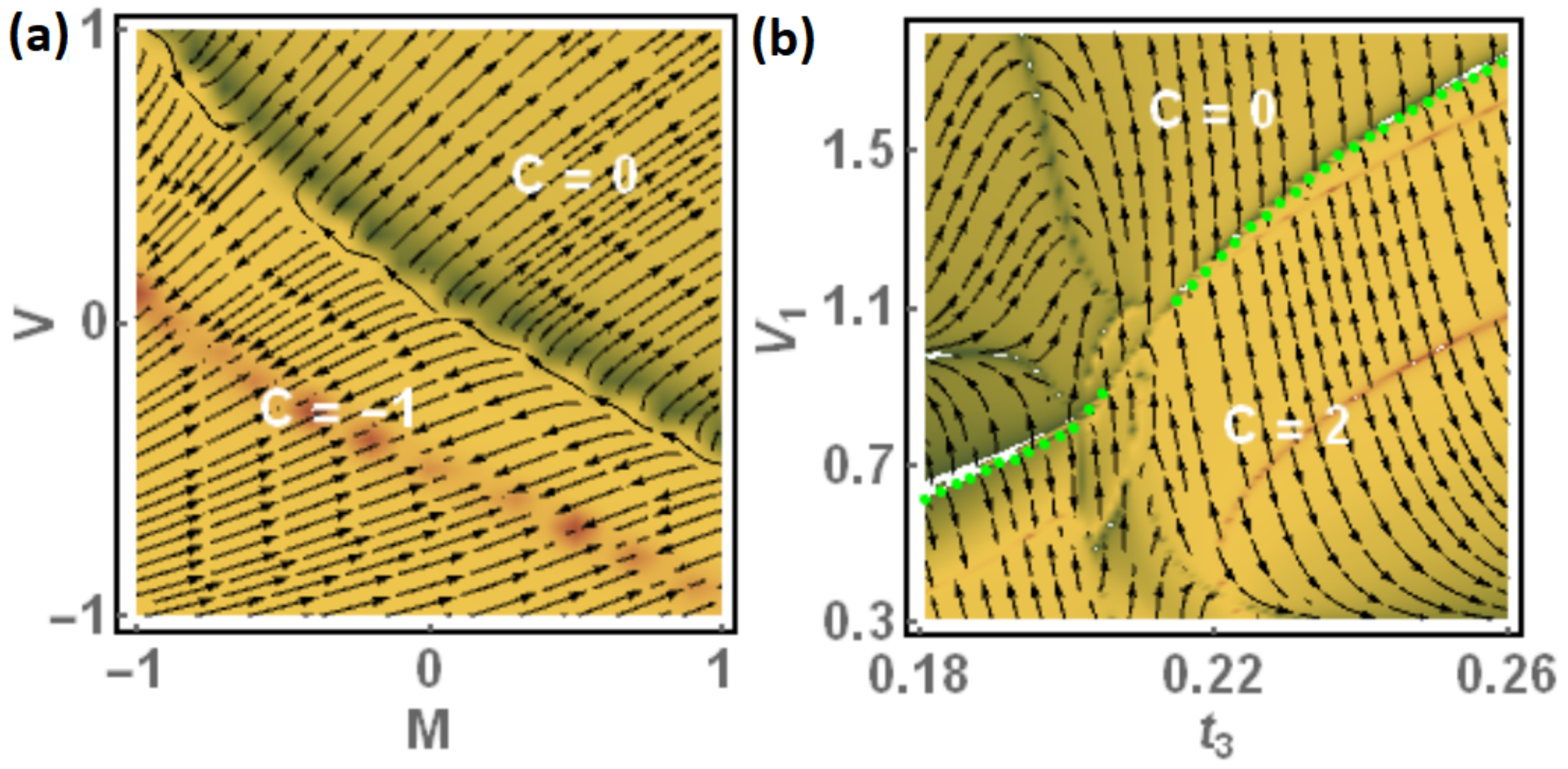}
\caption{(2) RG flow of a 2D Chern insulator in the presence of weak nearest-neighbor interaction in the parameter space ${\bf M}=(M,V)$\cite{Chen:2018}, solved by means of Green's function, and (b) that of the triangular lattice model of FCI in the parameter space ${\bf M}=(t_{3},V_{1})$, solved by means of twisted boundary condition and exact diagonalization. Both models extract critical exponents $\gamma=2$ and $\nu=1$ with respect to either the noninteracing or interacting parameter.}
\label{fig:interacting_results}
\end{figure}

\section{Weakly interacting systems}
We now briefly discuss how to incorporate the effects of interaction into our CRG formalism.
In weakly interacting systems,  interactions can be  perturbatively  treated using the Matsubara Green's function $G_{IJ}({\bf k},\sigma)=-\langle T_{\sigma}c_{I{\bf k}}(\sigma)c_{J{\bf k}}^{\dag}(0)\rangle$ and Dyson's equation $G=G_{0}+G_{0}\Sigma G$, where $\Sigma$ is the self-energy. 
The topological invariant is then calculated from the integration of specific combinations of the Green's functions which depend on the dimension and symmetry class of the system\cite{Niu1985a,Essin11,Gurarie11}. 
Consequently, the Fourier transform of the curvature function represents the amplitude of the convoluted Green's function propagating over a certain distance\cite{Chen:2018}. 
%Another remarkable advantage of this Green's function approach is that
Moreover, the gap closure at the TPTs can be identified from the spectral function, and the zero energy topological edge state can be pinpointed from the local density of states, corroborating the bulk-edge correspondence in the presence of interactions\cite{Zegarra19}.

%As a concrete example, 
In 2D systems that break time reversal symmetry and have a frequency-independent self-energy, the curvature function takes the same form as Eq.~(\ref{Berry_curvature_2D}), except the ${\bf d}$-vector is replaced by the self-energy-renormalized ${\bf d}^{\prime}={\bf d}-{\boldsymbol\Sigma}$. 
The ${\bf d}^{\prime}$-vector can be further written in terms of a linear combination of Green's functions denoted by $\tilde{G}({\bf k})$. 
The curvature function then takes the form\cite{Chen:2018}
\begin{eqnarray}
F({\bf k},{\bf M})=\frac{\pi}{i}\epsilon^{abc}\tilde{G}_{a}({\bf k})\partial_{x}\tilde{G}_{b}({\bf k})\partial_{y}\tilde{G}_{c}({\bf k})\;,
\end{eqnarray}
and can be shown to manifest the Lorentzian shape discussed in Eq.~(\ref{Lorentzian_shape}) despite interactions. 
The equivalent ``Wannier" correlation function reads %defined by the Fourier transform of the curvature function reads
\begin{eqnarray}
&&\tilde{F}({\bf R})=i\pi\int d^{2}{\bf R}_{1}\int d^{2}{\bf R}_{2}
\nonumber \\
&&\times\epsilon^{abc}\tilde{G}_{a}({\bf R}-{\bf R}_{1}-{\bf R}_{2})r_{1x}\tilde{G}_{b}({\bf R}_{1})r_{2y}\tilde{G}_{c}({\bf R})
\label{lambdaR_2D}
\end{eqnarray}
and represents the amplitude of the convoluted Green's function propagating over a certain distance, decaying with correlation length $\xi$. 
The 2D Chern insulator with electron-electron interaction calculated up to one-loop level is a concrete example for this case\cite{Chen:2018}, which features topological phase transitions driven by the mass term $M$ in the unperturbed Hamiltonian and the nearest-neighbor interaction $V$. 
The result of applying CRG to ${\bf k}_{0}=(0,0)$ in this model up to one-loop level is shown in Fig.~\ref{fig:interacting_results} (a). The RG flow identifies a phase boundary between the topologically nontrivial ${\cal C}=-1$ and trivial ${\cal C}=0$ phase in the ${\bf M}=(M,V)$ parameter space, which is a continuous line that passes through the critical point in the noninteracting limit\cite{Bernevig13} ${\bf M}_{c}=(0,0)$.
Furthermore, the extracted critical exponents $\gamma\approx 2$ and $\nu_{i}\approx 1$ reveal that the TPTs in the interacting system belong to the same universality class as the noninteracting 2D Chern insulators\cite{Chen:2017}.

\section{Strongly interacting systems}

For 2D systems with arbitrarily strong interaction, the topology can be implemented by imposing twisted periodic boundary conditions with phases ${\boldsymbol\phi}=(\phi_{1},\phi_{2})$ into the many-body Hamiltonian 
$\widehat{H}\equiv\widehat{H}({\boldsymbol \phi},{\bf M})$. In the thermodynamic limit $L^{\,}_1,L^{\,}_2\to\infty$ and on the torus (genus $g=1$),
the eigenenergies $E^{\,}_{m}(\bm{\phi},\bm{M})$
with $m=1,\cdots,N^{\,}_{g=1}$
are all degenerate, with the eigenstates denoted by $|\Psi^{\,}_{m}({\boldsymbol \phi},{\bf M})\rangle$. The Hall conductance at zero temperature $\sigma^{\,}_{\mathrm{H}}(\bf{M})$ is given by
\begin{equation}
\frac{\sigma^{\,}_{\mathrm{H}}(\bf{M})}{(e^{2}/h)}\equiv
\frac{1}{N^{\,}_{g=1}}
\sum_{n=1}^{N^{\,}_{g=1}}
C^{\,}_{m}(\bf{M}).
\label{eq: Niu Thouless formula a}
\end{equation}
%
%where the %integer-values
%
\begin{equation}
{\cal C}^{\,}_{m}({\bf M})\equiv
\lim_{L^{\,}_1,L^{\,}_2\to\infty}
\int\limits_{0}^{2\pi}\int\limits_{0}^{2\pi}
\frac{\mathrm{d}\phi^{\,}_{1}\,\mathrm{d}\phi^{\,}_{2}}{4\pi^2}
F^{\,}_{m}({\boldsymbol\phi},{\bf M})
\label{eq: Niu Thouless formula b}
\end{equation}
is the Chern number of the $m$-th state in the ground-state manifold and the domain of the integration is referred to as the flux Brillouin zone (fBZ).
The quantity
\begin{equation}
F^{\,}_{m}({\boldsymbol \phi},{\bf M})\equiv
4\pi\,
\mathrm{Im}
\left\langle\left.
\frac{
\partial\Psi^{\,}_{m}
     }
     {
\partial\phi^{\,}_{1}
     }
({\boldsymbol \phi},{\bf M})
\right|
\frac{
\partial\Psi^{\,}_{m}
     }
     {
\partial\phi^{\,}_{2}
     }
 ({\boldsymbol \phi},{\bf M})
\right\rangle,
\label{eq:berry-alt}
\end{equation}
is called the many-body Berry curvature~\cite{Thouless1982,Niu1985a} and is found to satisfy Eq.~(\ref{Lorentzian_shape}) in the flux space for the models that display linear many-body band crossing.
 The correlation function essentially takes the same form of Eq.~(\ref{Wannier_correlation_FT_Berry_curvature}), except the Wannier state %and position operators are 
is defined on the lattice dual to the fBZ, but once again decays with the correlation length $\xi$. 
Focusing on the case where only one $m$ of the $N^{\,}_{g=1}$ ground states in the thermodynamic limit has a non-vanishing ${\cal C}^{\,}_{m}({\bf M})$, an example of strongly-interacting topological system is the triangular-lattice model of fractional Chern insulator (FCI) at density $\rho=1/3$\cite{Venderbos2011a,Kourtis2012a,Kourtis2013a,Kourtis2013}. 
The CRG approach applied to this model yields the RG flow shown in Fig.~\ref{fig:interacting_results} (b)\cite{Kourtis:2017}. 
Despite the strong interaction, the model extracts critical exponents $\nu=1$ and $\gamma=2$ with respect to either the noninteracting or interacting tuning parameters, indicating the same universality class as the noninteracting limit. Moreover, in a variety of FCI models, the jump of topological invariant $\Delta{\cal C}$ across the TPT is found to be always equal to the sum of the order of band crossing at each gap-closing point\cite{Kourtis:2017}
\begin{eqnarray}
\Delta{\cal C}=\sum_{i=1}^{N_{div}}n_{i}\;,
\end{eqnarray}
exactly as in the noninteracting cases\cite{Chen-Schnyder:2019}. 
Finally, note that various interacting models have been shown to exhibit first order TPTs\cite{Budich13,Amaricci15,Roy16,Imriska16,Amaricci17,Barbarino19}. Although further investigations are necessary to address them, based on how transitions are characterized in statistical mechanics, we anticipate that first order topological transitions require a different paradigm than the critical behavior discussed in the present work.  An exciting  future direction would be the development of an unifying framework describing both first and second order topological phase transitions.

\section{Conclusions and Outlook}
Based on the %simple observation of a 
divergent behavior of topological curvature functions, we developed a unified description for quantum criticality near TPTs that is applicable to a broad range of noninteracting, weakly interacting, strongly interacting, and periodically driven topological systems.
The CRG approach has proved to be highly efficient in solving an array of models in a multi-dimensional parameter space.
For both static and driven noninteracting gapped systems, the critical exponents and scaling laws permit the classification of TPTs into universality classes which are determined by the underlying nonspatial and crystalline symmetries.
%For driven systems, we further uncover the possibility of generating quantum multi-criticality and engineering exotic topological phases such as nodal loop semimetals
%This is connected to emergent symmetries which may also feature in other Floquet topological systems.
For weakly interacting models, using Green's functions calculations up to one-loop shows that the system remains in the same universality class as the non-interacting problem.
It remains to be verified if higher loop corrections preserve this feature.
In the strongly interacting models investigated by means of twisted boundary conditions, the TPTs can still be classified into the same universality classes as the noninteracting theories despite the lack of a simple low energy Dirac form. 

These approaches can be straightforwardly generalized to higher dimensions and other symmetry classes. 
Beyond a classification of TPTs, we expect critical exponents to play important roles in the dynamics of topological systems.
An example would be characterizing the dynamics when the system is quenched across a TPT.  
Generalizing the formalism presented here to gapless materials, such as topological semimetals, as well as higher order topological insulators and especially non-hermitian systems, are natural directions for future explorations. 
It would also be interesting to incorporate finite temperature effects to see if TPTs survive and whether one can develop the notion of topological quantum critical scaling at finite temperatures.

%\acknowledgments
%Insert here the text.
\bibliographystyle{unsrt}

\end{document}